\documentclass{bioinfo}
\copyrightyear{2021}
\pubyear{2021}

\usepackage{graphicx}
\usepackage{hyperref}
\usepackage{url}
\usepackage{amsmath}
\usepackage[ruled,vlined]{algorithm2e}

\SetCommentSty{mycommfont}
\SetKwComment{Comment}{$\triangleright$\ }{}

\usepackage{natbib}
\begin{document}
\firstpage{1}

\title[Improvements to minimap2]{New strategies to improve minimap2 alignment accuracy}
\author[Li]{Heng Li$^{1,2}$}
\address{$^1$Dana-Farber Cancer Institute, 450 Brookline Ave, Boston, MA 02215, USA,
$^2$Harvard Medical School, 10 Shattuck St, Boston, MA 02215, USA}

\maketitle

\begin{abstract}

\section{Summary:} We present several recent improvements to minimap2, a
versatile pairwise aligner for nucleotide sequences. Now minimap2 v2.22 can
more accurately map long reads to highly repetitive regions and align through
insertions or deletions up to 100kb by default, addressing major weakness in
minimap2 v2.18 or earlier.

\section{Availability and implementation:}
\href{https://github.com/lh3/minimap2}{https://github.com/lh3/minimap2}

\section{Contact:} hli@ds.dfci.harvard.edu
\end{abstract}

\section{Introduction}
Minimap2~\citep{Li:2018ab} is widely used for maping long sequence
reads and assembly contigs. \citet{Jain:2020aa} found minimap2 v2.18 or earlier occasionally
misaligned reads from highly repetitive regions as minimap2 ignored seeds of
high occurrence. They also noticed minimap2 may misplace reads with structural
variations (SVs) in such regions~\citep{Jain2020.11.01.363887}. These
misalignments have become a pressing issue in the advent of
temolere-to-telomore human assembly~\citep{Miga:2020aa}. Meanwhile, old minimap2
was unable to efficiently align long insertions/deletions (INDELs) and often
breaks an alignment around variable-number tandem repeats (VNTRs). This has
inspired new chaining algorithms~\citep{Li:2020aa,Ren:2021aa} which are not
integrated into minimap2. Here we will describe recent efforts implemented
in v2.19 through v2.22 to improve mapping results.

\begin{methods}
\section{Methods}

\subsection{Rescuing high-occurrence $k$-mers}
Minimap2 keeps all $k$-mer minimizers~\citep{Roberts:2004fv} during indexing. Its original
implementation only selected low-occurrence minimizers during mapping. The
cutoff is a few hundred for mapping long reads against a human genome. If a
read habors only a few or even no low-occurrence minimizers, it will fail
chaining due to insufficient anchors.

To resolve this issue, we implemented a new heuristic to add additional
minimizers. Suppose we are looking at two adjacent low-occurence $k$-mers
located at position $x_1$ and $x_2$, respectively. If $|x_1-x_2|\ge500$,
minimap2 v2.22 additionally selects $\lfloor|x_1-x_2|/500\rfloor$ minimizers
of the lowest occurrence among minimizers between $x_1$ and $x_2$.
We use a binary heap data
structure to select minimizers of the lowest occurrence in this interval.
This strategy adds necessary anchors at the cost of increasing total alignment
time by a few percent on real data.

\subsection{Aligning through longer INDELs}
The original minimap2 may fail to align long INDELs due to its chaining
heuristics. Briefly, minimap2 applies dynamic programming (DP) to chain
minimizer anchors. This is a quadratic algorithm, slow for chaining
contigs. For acceptable performance, the original minimap2 uses a 500bp band by
default, which means a gap longer than 500bp will stop chaining.
To align through longer gaps, older minimap2 implemented a long-join heurstic as follows.
If there is an INDEL longer than 500bp and the two chains around the INDEL
have no overlaps on either the query or the reference sequence, minimap2 may
join the two short chains later.
This heuristic may fail around VNTRs because short chains
often have overlaps in VNTRs. More subtly, minimap2 may escape the inner DP
loop early, again for performance, if the chaining result is not improved for
50 iterations. When there is a copy number change in a long segmental
duplication, the early escape may break around the event even if users
specify a large band.

In minigraph~\citep{Li:2020aa}, we developed a new chaining algorithm that
finds up to 1kb INDELs with DP-based chaining and goes through longer INDELs with a
subquadratic algorithm~\citep{DBLP:conf/wabi/AbouelhodaO03}. We ported the same
algorithm to minimap2 for contig mapping. For long-read mapping, the minigraph
algorithm is slower. Minimap2 v2.22 still uses the DP-based algorithm to
find short chains and then invokes the minigraph algorithm to rechain anchors in
these short chains. The rechaining step achieves the same goal as long-join
but is more reliable because it can resolve overlaps between short chains. The old
long-join heuristic has since been removed.

\subsection{Properly mapping long reads with SVs}
The original minimap2 ranks an alignment by its Smith-Waterman score and
outputs the best scoring alignment. However, when there are SVs on the read,
the best scoring alignment is sometimes not the correct alignment.
\citet{Jain2020.11.01.363887} resolved this dilemma by altering the mapping
algorithm.

In our view, this problem is rooted in impropriate scoring: affine-gap penalty
over-penalizes a long INDEL that was often evolutionarily created in one event.
We should not penalize a SV by a function linear in the SV length. Minimap2 v2.22 instead rescores
an alignment with the following scoring function. Suppose an alignment consists
of $M$ matching bases, $N$ substitutions and $G$ gap opens, we empirically
score the alignment with
$$
S=M-\frac{N+G}{2d}-\sum_{i=1}^G\log_2(1+g_i)
$$
where $g_i\ge1$ is the length of the $i$-th gap and
$$
d=\max\left\{\frac{N+G}{M+N+G},0.02\right\}
$$
It approximates per-base sequence divergence except with the smallest value set
to 2\%. As an analogy to affine-gap scoring, the matching score in our scheme
is 1, the mismatch and gap open penalties are both $1/2d$ and the gap extension
penalty is a logarithm function of the gap length. Our scoring gives a long SV
a much milder penalty. In terms of time complexity, scoring an alignment is
linear in the length of the alignment. The time spent on rescoring is negligible in
practice.


\end{methods}

\section{Results}

\begin{table}
\processtable{Evaluation of minimap2 v2.22}
{\footnotesize\label{tab:1}\begin{tabular}{p{4.2cm}rrrr}
\toprule
$[$Benchmark$]$ Metric & v2.22 & v2.18 & Winno & lra \\
\midrule
$[$sim-map$]$ \% mapped reads at Q10      & 97.9 & 97.6 & {\bf 99.0} & 97.3 \\
$[$sim-map$]$ err. rate at Q10 (phredQ)   & {\bf 52}   & {\bf 52}   & 38   & 24 \\
$[$winno-cmp$]$  rate of diff. (phredQ)   & {\bf 41}   & 37   & N/A  & 18 \\
$[$sim-sv$]$  \% false negative rate      & {\bf 0.5}  & 2.0  & {\bf 0.5}  & 1.4  \\
$[$sim-sv$]$  \% false discovery rate     & {\bf 0.0}  & 0.1  & {\bf 0.0}  & 0.1  \\
$[$real-sv-1k$]$ \% false negative rate   & {\bf 7.3}  & 20.0 & 13.0 & N/A \\
$[$real-sv-1k$]$ \% false discovery rate  & 2.7  & {\bf 2.4}  & 2.7  & N/A \\
\botrule
\end{tabular}}
{In $[$sim-map$]$, 152,713 reads were simulated from the CHM13 telomere-to-telomere assembly v1.1
(AC: GCA\_009914755.3) with pbsim2~\citep{Ono:2021aa}: ``pbsim2 -{}-hmm\_model R94.model -{}-length-min
5000 -{}-length-mean 20000 -{}-accuracy-mean 0.95''. Alignments of mapping quality
10 or higher were evaluated by ``paftools.js mapeval''. The mapping error rate
is measured in the phred scale: if the error rate is $e$, $-10\log_{10}e$ is
reported in the table. In $[$winno-cmp$]$, 1.39 million CHM13 HiFi reads from
SRR11292121 were mapped against the same CHM13 assembly. 99.3\% of them were mapped by Winnowmap2
at mapping quality 10 or higher and were taken as ground truth to evaluate
minimap2 and lra with ``paftools.js pafcmp''. $[$sim-sv$]$ simulated 1,000
50bp to 1000bp INDELs from chr8 in CHM13 using SURVIVOR~\citep{Jeffares:2017aa} and simulated Nanopore
reads at 30-fold coverage with the same pbsim2 command line. SVs were called with
``sniffles -q 10''~\citep{Sedlazeck:2018ab} and compared to the simulated truth with ``SURVIVOR eval
call.vcf truth.bed 50''. In $[$real-sv-1k$]$, small and long variants were
called by dipcall-0.3~\citep{Li:2018aa} for HG002 assemblies (AC: GCA\_018852605.1 and
GCA\_018852615.1) and compared to the GIAB truth~\citep{Zook:2020aa} using ``truvari -r 2000 -s
1000 -S 400 -{}-multimatch -{}-passonly'' which sets the minimum INDEL size to 1kb in evaluation. }
\end{table}

We evaluated minimap2 v2.22 along with v2.18, Winnowmap2 v2.03 and lra v1.3.2
(Table~\ref{tab:1}). Both versions of minimap2 achieved high mapping accuracy on
simulated Nanopore reads (sim-map). Winnowmap2 aligned more reads at mapping
quality 10 or higher (mapQ10). However, it may occasionally assign a high mapping
quality to a read with multiple identical best alignments. This reduced its
mapping accuracy.

In lack of groud truth for real data, we took Winnowmap2 mapping as ground
truth to evaluate other mappers (winno-cmp in Table~\ref{tab:1}). Out of 1,378,092 reads with mapQ10
alignments by Winnowmap2, minimap2 v2.22 could map all of them. 118 reads, less
than 0.01\% of all reads, were mapped differently by v2.22. 51 of them have
multiple identical best alignments. We believe these are more likely to be
Winnowmap2 errors. Most of the remaining 67 (=118-51) reads have multiple
highly similar but not identical alignments. We are not sure how many are real
mapping errors. Minimap2 v2.18 is less consistent with Winnowmap2. Most of the
differences are located in highly repetitive regions.

The two benchmarks above only evaluate read mappings when there are no variations between the reads and the reference.
To measure the mapping accuracy in the presence of SVs (sim-sv), we reproduced
the results by~\citep{Jain2020.11.01.363887}. Minimap2 v2.22 is as good as
Winnowmap2 now. Note that we were setting the Sniffles mapping quality
threshold to 10 in consistent with the benchmarks above. If we used the
default threshold 20, v2.22 would miss additional five SVs (accounting for
0.5\% of simulated SVs). For four out of these missing five SVs, minimap2 v2.22
mapped more variant reads than Winnowmap2. Sniffles did not call these SVs
because minimap2 tended to give them conservative mapping quality. It is worth
noting that the simulation here only considers a simple scenario in evolution.
Non-allelic gene conversions, which happen often in segmental
duplications~\citep{Harpak:2017aa}, would obscure the optimal mapping
strategies. How much such simple SV simulation informs real-world SV calling
remains a question.

To see if minimap2 v2.22 could improve long INDEL alignment, we ran dipcall on
contig-to-reference alignments and focused on INDELs longer than 1kb
(real-sv-1k). v2.22 is more sensitive at comparable specificity, confirming its
advantage in more contiguous alignment. lra is supposed to handle long INDELs
well, too. However, we could not get dipcall to work well with lra,
so did not report the numbers.

Minimap2 spends most computing time on base alignment. As recent improvements
in v2.22 incur little additional computing and do not change the base alignment
algorithm, the new version has similar performance to older verions. It is
consistently faster than Winnowmap2 by several times. Sometimes simple
heuristics can be as effective as more sophisticated yet slower solutions.

\section*{Acknowledgements}
We thank Arang Rhie and Chirag Jain for providing motivating examples for which
older minimap2 underperforms.

\paragraph{Funding\textcolon} This work is funded by NHGRI grant R01HG010040.

\bibliography{minimap2}

\begin{thebibliography}{}

\bibitem[Abouelhoda and Ohlebusch, 2003]{DBLP:conf/wabi/AbouelhodaO03}
Abouelhoda, M.~I. and Ohlebusch, E. (2003).
\newblock A local chaining algorithm and its applications in comparative
  genomics.
\newblock In {\em Algorithms in Bioinformatics, Third International Workshop,
  {WABI} 2003, Budapest, Hungary, September 15-20, 2003, Proceedings}, pages
  1--16.

\bibitem[Harpak et~al., 2017]{Harpak:2017aa}
Harpak, A. et~al. (2017).
\newblock Frequent nonallelic gene conversion on the human lineage and its
  effect on the divergence of gene duplicates.
\newblock {\em Proc Natl Acad Sci U S A}, 114:12779--12784.

\bibitem[Jain et~al., 2020a]{Jain2020.11.01.363887}
Jain, C. et~al. (2020a).
\newblock A long read mapping method for highly repetitive reference sequences.
\newblock {\em bioRxiv}.

\bibitem[Jain et~al., 2020b]{Jain:2020aa}
Jain, C. et~al. (2020b).
\newblock Weighted minimizer sampling improves long read mapping.
\newblock {\em Bioinformatics}, 36:i111--i118.

\bibitem[Jeffares et~al., 2017]{Jeffares:2017aa}
Jeffares, D.~C. et~al. (2017).
\newblock Transient structural variations have strong effects on quantitative
  traits and reproductive isolation in fission yeast.
\newblock {\em Nat Commun}, 8:14061.

\bibitem[Li, 2018]{Li:2018ab}
Li, H. (2018).
\newblock Minimap2: pairwise alignment for nucleotide sequences.
\newblock {\em Bioinformatics}, 34:3094--3100.

\bibitem[Li et~al., 2018]{Li:2018aa}
Li, H. et~al. (2018).
\newblock A synthetic-diploid benchmark for accurate variant-calling
  evaluation.
\newblock {\em Nat Methods}, 15(8):595--597.

\bibitem[Li et~al., 2020]{Li:2020aa}
Li, H. et~al. (2020).
\newblock The design and construction of reference pangenome graphs with
  minigraph.
\newblock {\em Genome Biol}, 21:265.

\bibitem[Miga et~al., 2020]{Miga:2020aa}
Miga, K.~H. et~al. (2020).
\newblock Telomere-to-telomere assembly of a complete human {X} chromosome.
\newblock {\em Nature}, 585:79--84.

\bibitem[Ono et~al., 2021]{Ono:2021aa}
Ono, Y. et~al. (2021).
\newblock {PBSIM2}: a simulator for long-read sequencers with a novel
  generative model of quality scores.
\newblock {\em Bioinformatics}, 37:589--595.

\bibitem[Ren and Chaisson, 2021]{Ren:2021aa}
Ren, J. and Chaisson, M. J.~P. (2021).
\newblock lra: A long read aligner for sequences and contigs.
\newblock {\em PLoS Comput Biol}, 17:e1009078.

\bibitem[Roberts et~al., 2004]{Roberts:2004fv}
Roberts, M. et~al. (2004).
\newblock Reducing storage requirements for biological sequence comparison.
\newblock {\em Bioinformatics}, 20:3363--9.

\bibitem[Sedlazeck et~al., 2018]{Sedlazeck:2018ab}
Sedlazeck, F.~J. et~al. (2018).
\newblock Accurate detection of complex structural variations using
  single-molecule sequencing.
\newblock {\em Nat Methods}, 15:461--468.

\bibitem[Zook et~al., 2020]{Zook:2020aa}
Zook, J.~M. et~al. (2020).
\newblock A robust benchmark for detection of germline large deletions and
  insertions.
\newblock {\em Nat Biotechnol}, 38:1347--1355.

\end{thebibliography}

\end{document}